\documentclass[12pt]{iopart}
\usepackage{orcidlink}
\usepackage{hyperref}
\usepackage{float}
\usepackage{hhline}
\usepackage{multirow}
\usepackage{ragged2e}
\usepackage{epsfig}
\usepackage{caption}
\usepackage{latexsym}
\usepackage{bm}
\usepackage{amsfonts}
\usepackage[latin1]{inputenc}
\usepackage{iopams}
\usepackage{graphicx}
\usepackage{palatino}
\usepackage{mathpazo}
\usepackage{hhline}
\usepackage{multirow}
\usepackage{textcomp}
\usepackage{booktabs}
\usepackage{hhline}
\usepackage{xcolor}
\hypersetup{colorlinks,citecolor=blue}
\hypersetup{colorlinks=true,linkcolor=red,filecolor=magenta, urlcolor=blue}

\begin{document}

\title[Preparing for IOP Publishing journals]{Observational constrained $F(R,\mathcal{G})$ gravity cosmological model and the dynamical system analysis}

\author{Santosh V. Lohakare$^{1,a}$\orcidlink{0000-0001-5934-3428}, Krishna Rathore$^{2,a}$\orcidlink{0009-0009-1158-0569} and B. Mishra$^{3,a,*}$\orcidlink{0000-0001-5527-3565}}

\address{$^a$ Department of Mathematics, Birla Institute of Technology and Science-Pilani, Hyderabad Campus, Hyderabad-500078, India.}
\ead{$^1$lohakaresv@gmail.com, $^2$rathorekrishnaa2000@gmail.com and $^3$bivudutta@yahoo.com}
\vspace{10pt}

\begin{abstract}
In this paper, we have analyzed the geometrical and dynamical parameters of $\mathcal{F}(R, \mathcal{G})=\alpha R^2 \mathcal{G}^\beta$ cosmological model, ($R$, $\mathcal{G}$ being the Ricci scalar and Gauss-Bonnet invariant respectively), constraining the parameters through the cosmological data sets. It is exhibited that the model admits a viable radiation era, and early deceleration followed by late-time acceleration in the matter-dominated era. From the phase-space, portrait stability criterion has been analysed, restricting the parameter $\beta$, different from $\beta=-1$. Additionally, we have explored the stability of the model from the behavior of critical points and obtained the present value of the density parameter for matter-dominated and dark energy components, which are identical to those obtained through cosmological data sets.
\end{abstract}

\textbf{Keywords}: Gauss-Bonnet invariant, Cosmological data set, Dynamical system analysis, Critical points.
\vspace{10pt}

* Author to whom any correspondence should be addressed.
%
%
%
%
%

\section{Introduction} \label{SEC-I}
The issue of late-time cosmic acceleration of the Universe is one of the prominent cosmological problems in modern cosmology. An exotic form of energy known as dark energy (DE) is considered to be responsible for this strange behavior of the Universe. High-precision observational data from Type Ia supernovae (SNe Ia) \cite{Riess_1998_116, Perlmutter_1998_517},  Wilkinson Microwave Anisotropy Probe (WMAP) experiment \cite{Spergel_2003_148}, Baryonic Acoustic Oscillations (BAO) \cite{Eisenstein_2005_633}, Cosmic Microwave Background (CMB) \cite{Hinshaw_2013_208}, Baryon Oscillation Spectroscopic Survey (BOSS) \cite{Alam_2017_470}, Large-Scale Structure (LSS) \cite{Daniel_2008_77} and Planck Collaboration \cite{Ade_2016_594a, Aghanim_2018_641} have strongly supported this behavior. The limitation in addressing this issue under the purview of General Relativity (GR) compels its modification. Different exotic matter fields that simulates negative pressure and positive energy density can be used as an additional term to settle this issue. However, it can also be settled through a suitable geometrical modification in GR, which resulted in modified theories of gravity. 

The Einstein field equations can be modified to fit the matter-energy content of the observable Universe by changing the geometrical sector. On the explanation of the evolution of the Universe, various modified gravity theories have been proposed \cite{Capozziello_2011_509, Faraoni_2010_170, Nojiri_2011_505, Carroll_2004_70, Nojiri_2007_04, Nojiri_2017_692}. One of the important findings is that it is possible to define early inflation with different coupling parameters and describe the late-time DE-dominated era with precision \cite{Starobinsky_1980_91}. In $f(R)$ gravity \cite{Carroll_2004_70, Nojiri_2011_505}, the gravitational action generalizes the Einstein-Hilbert action by introducing a generic function of the Ricci scalar curvature $R$ and GR can be restored by assuming $f(R) = R$. The general relativistic gravitational Lagrangian may be modified to include a broader range of curvature invariants, such as $R$, $R_{i\,j}R^{i\,j}$ and $R_{i\,j\,k\,l}R^{i\,j\,k\,l}$ among others. The $F(R, \mathcal{G})$ model acts as a viable alternative to dark energy \cite{Martino_2020_102, Benetti_2018_27, ODINTSOV_2019_938_935, Elizalde_2010_27}. The gravitational Lagrangian in Gauss-Bonnet (GB) gravity theories is a function $F(R, \mathcal{G})$, where the GB invariant $\mathcal{G}$ is defined as $\mathcal{G} \equiv R^2-4R^{i\,j} R_{i\,j}+R^{i\,j\,k\,l}R_{i\,j\,k\,l}$. In differential geometry and topology, the Gauss-Bonnet invariant modifies the Einstein-Hilbert action that governs the dynamics of gravity. In Refs. \cite{Baojiu_2007_76, Lattimer_2014_784, De_Felice_2009_675, Nojiri_2005_631, Cognola_2006_73}, the $F(R,\mathcal{G})$ gravity was proposed to incorporate $R$ and $\mathcal{G}$ into a bivariate function that supports the double inflationary scenario \cite{Laurentis_2015_91} and are also strongly supported by observations \cite{Capozziello_2014_29}. Besides its stability, the $F(R,\mathcal{G})$ theory is well-suited to describe the crossing of the phantom divide line and the transformation between an accelerating and decelerating state of celestial bodies.

In scalar-tensor gravity, phase space is vibrant due to the fourth-order contributions of the Gauss-Bonnet invariant and the second-order contributions of the scalar field \cite{Konstantinos_2022_2211.06076}. Several invariant structures in phase space are necessary for the theory to be valid and viable in describing the evolution of the Universe \cite{Chatzarakis_2020_419}. Using dynamical system analysis, Shah et al. \cite{Shah_2019_79} analyzed the stability properties and acceleration phase of the Universe under various circumstances. The combined study of the data $H(z)$ and $[f\sigma_8] (z)$ shows that for $n=2$, the Starobinsky model of $f(R)$ fits well with the observational data and is a feasible alternative to the $\Lambda$CDM model \cite{Bessa_2022_82}. Using the dynamical system approach and constraining observational data, Bayarsaikhan et al. \cite{Bayarsaikhan_2023_83} have examined regularized Einstein-Gauss-Bonnet gravity in four dimensions with a non-minimal scalar coupling function.

In cosmological observation, the cosmic chronometer (CC) approach can be used to determine the age and expansion rate of the Universe. The CC technique consists of three basic components: i) the definition of a sample of optimal CC tracers; ii) the determination of the differential age; and iii) the assessment of systematic effects \cite{Moresco_2022_25}. The value of the Hubble parameter $H(z)$ is instrumental in determining the energy content of the Universe and its acceleration mechanism. The estimation of $H(z)$ is carried out mainly at $z=0$. But there are methods to determine $H(z)$ such as the detection of BAO signal in the clustering of galaxies and quasars, analyzing SN data, Ref. \cite{Riess_2018_853, Riess_2021_908, Font_Ribera_2014_2014_027, Raichoor_2020_500, Hou_2020_500}) and so on. \textit{Pantheon}$^+$ is the successor to the original Pantheon analysis \cite{Scolnic_2018_859} and expands the original Pantheon analysis framework to combine an even larger number of SN Ia samples to understand the complete expansion history. Here, we have used the CC sample, \textit{Pantheon}$^+$, and BAO data sets to investigate the expansion history of the Universe, as well as the behavior of other geometrical parameters. 

Noether symmetry analysis revealed that one such symmetry is admissible for $F(R, \mathcal{G})=\alpha R^n \mathcal{G}^{1-n}$ \cite{Capozziello_2014_29}. Viable cosmological solutions that includes stability criteria for such a form have also been explored \cite{Santos_da_Costa_2018_35}. However, the absence of a linear term in the Ricci scalar is a serious concern, while other forms may also be possible from symmetry analysis. We consider $F(R, \mathcal{G})=R + \alpha R^2 \mathcal{G}^\beta$, keeping $\alpha$ and $\beta$ to be arbitrary constants, different from $\beta = -1$ and study the cosmological consequence and the stability criteria \cite{Santos_da_Costa_2018_35}. The article is organized as follows: In Section \ref{SEC-II}, we present the mathematical formalism of $F(R, \mathcal{G})$ gravity. Section \ref{SEC-III} discusses and uses the observational data sets derived from the CC sample, \textit{Pantheon$^+$} samples and the BAO. The geometrical and dynamical parameters are also constrained by using these data sets.  Dynamical system analysis has been performed for the model in Section \ref{SEC-IV}. Finally, we summarize our results in Section \ref{SEC-V} with the conclusion.


\section{Basic Formalism of \texorpdfstring{$F(R, \mathcal{G})$}{} Gravity and Cosmology} \label{SEC-II}

The action of $F(R, \mathcal{G})$ gravity, a modification of GR \cite{Laurentis_2015_91, Wu_2015_92, Santos_da_Costa_2018_35, ODINTSOV_2019_938_935, Kumar_Sanyal_2020_37, Brout_2022_938} is,
\begin{equation}\label{1}
S=\int \sqrt{-g}\left[\frac{1}{2k^2}F(R,\mathcal{G})+\mathcal{L}_m\right] d^{4}x,
\end{equation}
where $g$ is a metric determinant, $\mathcal{L}_m$ describes Lagrangian matter, $k^2=8\pi G_N$, $G_N$ is the gravitational constant. The Gauss-Bonnet invariant is defined as
\begin{equation}\label{2}
\mathcal{G} \equiv R^2-4 R^{i\,j} R_{i\,j} + R^{i\,j\,k\,l} R_{i\,j\,k\,l},
\end{equation}
with the Ricci tensor and Riemann tensor, respectively, denoted by $R^{i\,j}$ and $R^{i\,j\,k\,l}$. The definition of $\mathcal{G}$ in differential geometry is
\begin{equation}\label{3}
\int_{\mathcal{M}} \mathcal{G} d^{n}x=\chi (\mathcal{M}),
\end{equation}
in 4-D, $\mathcal{G}=R^{i\,j\,k\,l} R_{i\,j\,k\,l}=\chi (\mathcal{M})$, which is metric independent, and so a topologically invariant Euler number. Consequently, $\int \mathcal{G} \sqrt{-g} d^{4}x $ yields a surface term. Thus GB term contributes only either through dynamical coupling or considering non-linear term. Here, we consider curvature scalar coupled non-linear GB term. By varying the action (\ref{1}) with respect to the metric tensor $g_{i\,j}$, the field equations of $F(R,\mathcal{G})$ gravity can be written as,
\begin{eqnarray} \label{4}
\nonumber F_R{G}_{i\,j}&=&k^2{T}_{i\,j}+\frac{1}{2}g_{i\,j}[F(R,\mathcal{G})-R F_{R}]+\nabla_{i} \nabla_{j} F_{R} - g_{i\,j} \Box F_{R} + F_{\mathcal{G}}\Big({-2R}{R_{i\,j}} \nonumber\\& &+4R_{i\,k}R^{k}_{j}-2R^{k\,l\,m}_{i}R_{j\,k\,l m} + 4g^{k\,l} g^{m\,n} R_{i\,k\,j\,m} R_{l\,n}\Big) +2(\nabla_{i}\nabla_{j}F_{\mathcal{G}})R \nonumber\\& &- 2g_{i\,j}(\Box F_{\mathcal{G}})R + 4(\Box F_{\mathcal{G}})R_{i\,j}-4(\nabla_{k} \nabla_{i} F_{\mathcal{G}})R^{k}_{j}-4(\nabla_{k} \nabla_{j} F_{\mathcal{G}})R^{k}_{i} \nonumber\\& &+ 4g_{i\,j}(\nabla_{k} \nabla_{l} F_{\mathcal{G}})R^{kl}-4(\nabla_{l} \nabla_{n} F_{\mathcal{G}})g^{kl}g^{mn}R_{i\,k\,j\,m}\, ,
\end{eqnarray}
where $G_{i\,j}$ represents the Einstein tensor, $\nabla_{i}$ describes the covariant derivative operator associated with $g_{i\,j}$, $\Box \equiv g^{i\,j}\nabla_{i}\nabla_{j}$ represents the covariant d'Alembert operator, and ${T}_{i\,j}$ represents the energy-momentum tensor. Additionally, the following quantities have been specified.
\begin{equation*} 
F_R\equiv \frac{\partial F(R,\mathcal{G})}{\partial R},\hspace{1cm} F_\mathcal{G}\equiv \frac{\partial F(R,\mathcal{G})}{\partial \mathcal{G}}.
\end{equation*}

The spacetime for the flat FLRW metric can be given as
\begin{equation} \label{5}
ds^{2}=-dt^{2}+a^{2}(t)(dx^{2}+dy^{2}+dz^{2}),
\end{equation}
where $a(t)$ is the scale factor and the Hubble parameter, $H\equiv\frac{\dot{a}(t)}{a(t)}$. The over-dot denotes the derivative with respect to cosmic time $t$. Subsequently, the Ricci scalar and the Gauss-Bonnet invariant respectively, becomes

\begin{equation} \label{6}
R=6(\dot{H}+2H^{2}), \hspace{1cm} \mathcal{G}=24H^{2}(\dot{H}+H^{2})
\end{equation}

Using an energy-momentum tensor, the Einstein equations and continuity equation are determined by the presence of an isotropic perfect fluid
\begin{eqnarray} \label{7}
    T_{i}^{j}=diag (-\rho, p, p, p),
\end{eqnarray}
where $\rho$ denotes the matter-energy density and $p$ is the pressure of matter.
By substituting equation (\ref{5}) and (\ref{6}) into the gravitational field equation (\ref{4}), we obtain the field equations of $F(R,\mathcal{G})$ gravity as,

\begin{eqnarray} 
3H^{2} F_{R}&=&{\kappa^{2}} \rho+\frac{1}{2}\left[R {F_{R}}+\mathcal{G} {F_{\mathcal{G}}} - {F(R,\mathcal{G})}\right]-3H \dot{F}_{R} -12H^{3} \dot{F}_{\mathcal{G}},\label{8}\\
(2\dot{H}+3H^{2}) F_{R} &=& -\kappa^{2} p +\frac{1}{2}\left[R F_{R}+\mathcal{G} F_{\mathcal{G}} - F(R,\mathcal{G})\right] -2H\dot{F}_{R}-\ddot{F}_{R}\nonumber\\& &-8H\left(\dot{H}+H^2\right) \dot{F}_{\mathcal{G}}-4H^{2} \ddot{F}_{\mathcal{G}}.\label{9}
\end{eqnarray}

Background cosmology can be simplified by rewriting these equations as effective fluids, embodying additional terms due to higher-order curvature terms incorporated into the expression. We consider the mapping, $F(R,\mathcal{G}) \longrightarrow R+\mathcal{F}(R,\mathcal{G})$ \cite{Marco_2020_135}. The motivation behind considering this form is its consistency with the concordance $\Lambda$CDM model. For  $\mathcal{F}=-2\Lambda$, with $\Lambda$ being the cosmological constant, it corresponds to $\Lambda$CDM paradigm. Accordingly, equation (\ref{8}) and (\ref{9}) reduce to,

\begin{eqnarray}
    3 H^2 &=& \kappa^2 (\rho_m+\rho_r + \rho_{DE}) = \kappa^2 \rho_{eff}, \label{10}\\
    3 H^2+2 \dot{H} &=& -\kappa^2 (p_m+p_r+p_{DE}) = - \kappa^2 p_{eff}, \label{11}
\end{eqnarray}
and resulting in the following identities as
\begin{eqnarray}
    \kappa^2 \rho_{DE} &=& -3 H^2 \mathcal{F}_R+\frac{1}{2}\big(R \mathcal{F}_R + \mathcal{G} \mathcal{F}_\mathcal{G}-\mathcal{F}(R,\mathcal{G}) - 6 H \dot{\mathcal{F}}_R-24 H^3 \dot{\mathcal{F}}_\mathcal{G}\big), \nonumber\\ \label{12}\\
    \kappa^2 p_{DE} &=& (2 \dot{H}+3 H^2) \mathcal{F}_R-\frac{1}{2} \big(R \mathcal{F}_R+\mathcal{G} \mathcal{F}_{\mathcal{G}}-\mathcal{F}(R,\mathcal{G}) - 4 H \dot{\mathcal{F}}_R\nonumber\\& & - 2\ddot{\mathcal{F}}_R-8 H^2 \ddot{\mathcal{F}}_{\mathcal{G}}-16 H \dot{H} \dot{\mathcal{F}}_{\mathcal{G}}-16 H^3 \dot{\mathcal{F}}_{\mathcal{G}}\big). \nonumber \label{13}\\
\end{eqnarray}

To solve the system [equations (\ref{12})- (\ref{13})], some viable form of $\mathcal{F}(R, \mathcal{G})$ would be required. Hence, we consider
\begin{eqnarray} \label{14}
\mathcal{F}(R,\mathcal{G})=\alpha R^2 \mathcal{G}^\beta ,
\end{eqnarray}
where $\alpha$ and $\beta$ are arbitrary constants, $\beta \neq 1$, and study cosmological consequence and the stability criteria, following the work \cite{Capozziello_2014_29}. It is a double inflationary scenario connected to the existence of Noether symmetries. The form of the function $\mathcal{F}(R, \mathcal{G})$ in modified gravity models can have intriguing cosmological implications. It can lead to modified field equations that govern the dynamics of the Universe, affecting the expansion rate, the evolution of cosmic structures, and the behavior of matter and energy. Studying the cosmological consequences of this form of $\mathcal{F}(R, \mathcal{G})$ allows researchers to explore new scenarios, such as inflationary models, DE models, and comparing them with the cosmological observations. In the case of a positive second derivative of $F(R, \mathcal{G})$ with respect to $R$, the model is free from instability within the context of Dolgov-Kawasaki instability \cite{Dolgov_2003_576}, and accordingly, the limits on the model parameters are $\alpha > 0$, and $\beta$ is even. We rewrite $R$ and $\mathcal{G}$ in the redshift parameter to get the expansion rate $\big[(1 + z)H(z) = -\frac{dz}{dt}\big]$ as
\begin{eqnarray}
    R&=&6\left(2 H_0^2 E(z)-\frac{H_0^2 (1+z) E'(z)}{2}\right), \nonumber\\ \mathcal{G}&=&24H_0^2 E(z) \left(H_0^2 E(z)-\frac{H_0^2 (1+z) E'(z)}{2}\right), \label{15}
\end{eqnarray}
where $H^2 (z)=H_0^2 E(z)$, $H_0$ represents the present value of the Hubble parameter, and the prime denotes the derivative to the redshift parameter. We use the following functional form for $E(z)$ \cite{Lemos_2018_483},
\begin{eqnarray} \label{16}
    E (z) = A\, (1+z)^3 + B + C\, z + D\, ln(1+z) ,
\end{eqnarray}
where $A$, $B$, $C$, and $D$ are free parameters. The $A(1+z)^3$ term accounts for the main effect of matter, as its energy density dilutes with the expansion of the Universe.  The terms $B$, $C\,z$  and $D\, ln(1+z)$ in the above expression are associated with the contribution from DE, which drives the accelerated expansion of the Universe. The flexible $E(z)$ allows us to explore alternative cosmological scenarios beyond standard $\Lambda$CDM model. Changing the parameters, one can examine the impact of different components and modifications of gravity on the expansion of the Universe.

\section{Observational Constraints} \label{SEC-III}
In cosmology, the Hubble and \textit{Pantheon$^+$} data sets are important to study the expansion history of the Universe and the properties of DE. Here, we shall use the early-type galaxies expansion rate data such as the $H(z)$, \textit{Pantheon$^+$} data, BAO and CMB distance priors. Since $H(z)$ provides the basic information about the energy content and the main physical mechanisms driving the present acceleration of the Universe; therefore the accurate determination of the expansion rate of the Universe has become important. In CC measurement, the expansion rate of the Universe is directly and cosmology-independently estimated without any assumptions about the origin of the Universe. There is no direct correlation between the observations and cosmological models. Therefore, these data sets serve as an independent tool to estimate the parameters of the cosmological models.

\begin{figure*} [!htb]
\centering
\includegraphics[width=120mm]{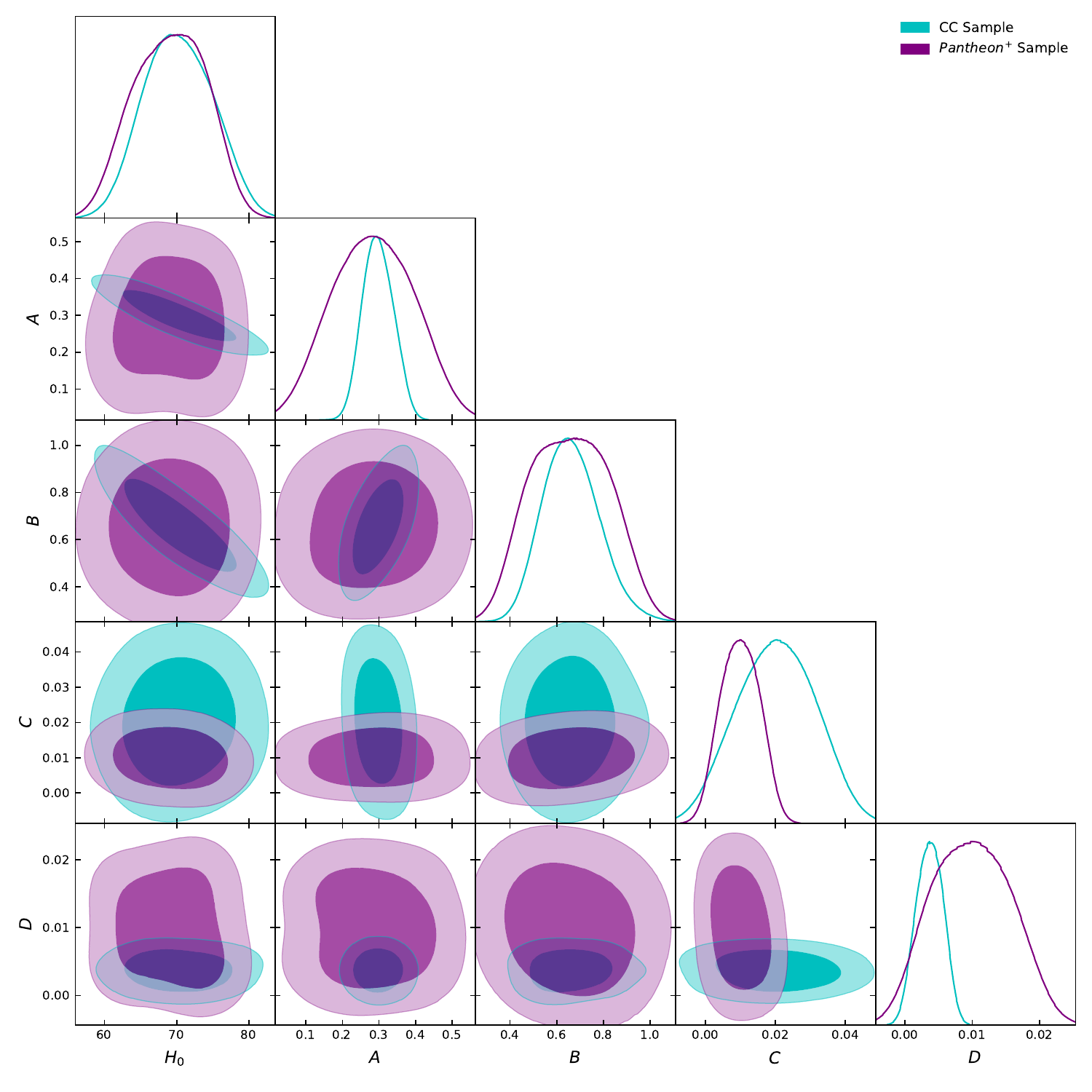}
\caption{The contour plots with $1-\sigma$ and $2-\sigma$ errors for the parameters $H_0$, $A$, $B$, $C$ and $D$ by using CC and {\textit{Pantheon$^+$}} datasets.}
\label{FIG2}
\end{figure*}

\begin{figure*} [!htb]
\centering
\includegraphics[width=120mm]{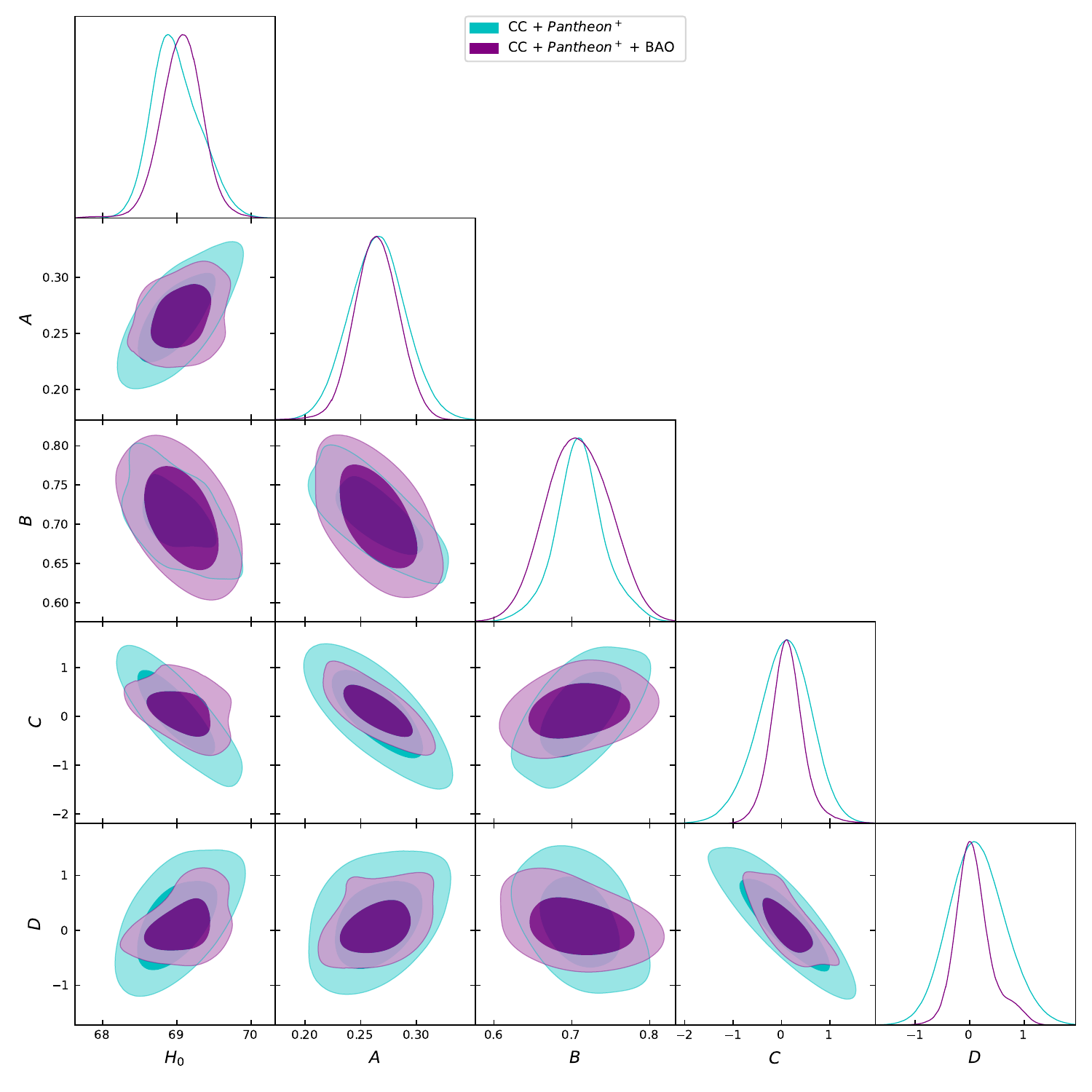}
\caption{The contour plots with $1-\sigma$ and $2-\sigma$ errors for the parameters $H_0$, $A$, $B$, $C$ and $D$ by using CC + {\textit{Pantheon$^+$}} + BAO datasets.}
\label{FIG3}
\end{figure*}

\begin{table*} [!htb]
\small\addtolength{\tabcolsep}{-2pt}
\centering
\begin{tabular}{|*{5}{c|}}\hline
\parbox[c][0.8cm]{2cm}{\textbf{Coefficients}} & \textbf{CC Sample}  & \textbf{\textit{Pantheon$^+$}} & \textbf{CC + \textit{Pantheon$^+$}} & CC + \textbf{\textit{Pantheon$^+$} + BAO}\\ [0.5ex]
\hline \hline
\parbox[c][0.7cm]{2cm}{$H_0$} & 70.2 $\pm$ 4.6 & 69.1 $\pm$ 4.8 & $68.69_{-0.59}^{+0.67}$ & $69.26_{-0.53}^{+0.57}$ \\
\hline
\parbox[c][0.7cm]{2cm}{$A$} & 0.297 $\pm$ 0.04 & 0.28 $\pm$ 0.11 & $0.285^{+0.050}_{-0.048}$ & $0.264^{+0.039}_{-0.036}$ \\
\hline
\parbox[c][0.7cm]{2cm}{$B$} & 0.66$^{+0.11}_{-0.13}$ & 0.64 $\pm$ 0.16 & $0.689^{+0.071}_{-0.067}$ & $0.698^{+0.070}_{-0.071}$ \\
\hline
\parbox[c][0.7cm]{2cm}{$C$} & 0.0099 $\pm$ 0.0053 & 0.02 $\pm$ 0.011 & $0.012^{+0.98}_{-1.1}$ & $0.012 \pm 0.71$ \\[0.5ex] 
\hline 
\parbox[c][0.7cm]{2cm}{$D$} & 0.0037 $\pm$ 0.0019 & 0.0099 $\pm$ 0.056 & $0.014^{+1.1}_{-0.99}$ & $0.0025^{+0.81}_{-0.61}$ \\[0.5ex]
\hline
\end{tabular}
\caption{These marginalized constraints are based on the CC, \textit{Pantheon$^+$} samples, and BAO data sets.}
\label{TABLE I}
\end{table*}

\subsection{CC Dataset}
To estimate the expansion rate of the Universe at redshift $z$, we use the widely used differential age (DA) method. In this way, it is possible to predict $H(z)$ using $(1+z) H(z)=-\frac{dz}{dt}$. The Hubble parameter is modeled on 32 data points (see {\bf APPENDIX}) for a redshift range of $0.07 \leq z \leq 1.965$ \cite{Moresco_2022_25}. The mean value of the parameters $H_0$, $A$, $B$, $C$ and $D$ are determined by minimizing the chi-square value. Using Hubble data, the chi-square function is as follows
\begin{equation} \label{eq.hubbdef}
\chi_{Hubble}^{2}=\sum_{i=1}^{32}\frac{\left[H_{th}(z_i)-H_{obs}(z_i)\right]^2}{\sigma_{i}^{'2}},
\end{equation}
where $z_i$ is the redshift at which $H(z_i)$ has been measured. A standard error in Hubble function experimental values is denoted by $\sigma_{i}^{'}$. The $H_{th}(z_i)$ and $H_{obs}(z_i)$ respectively indicate the theoretical and observable values of the Hubble parameter.

\subsection{\textit{Pantheon\texorpdfstring{$^+$}{}} Sample}
Among the \textit{Pantheon$^+$} sample data set are 1701 light curves of 1550 distinct Type Ia supernovae with redshifts between $(0.00122, 2.2613)$ \cite{Brout_2022_938}. The observed and theoretical distance moduli are compared to fit the model parameters. According to the \textit{Pantheon$^+$}, the SNe Ia functions for 1701 are
\begin{equation} \label{eq.pantheondef}
\chi_{\textit{Pantheon$^+$}}^{2}=\sum_{i=1}^{1701}\frac{\left[\mu_{th}(\mu_0,z_i)-\mu_{obs}(z_i)\right]^2}{\sigma_{i}^{'2}},
\end{equation}
where $\sigma_{i}^{'}$ is the standard deviation. The theoretical distance modulus $\mu_{th}$ is defined as $\mu_{th}^i=\mu(D_{L})=m-M=5 log_{10}D_L(z)+\mu_0$, where $m$ and $M$ are represented by apparent and absolute magnitudes, and the nuisance parameter $\mu_0$ is defined as $\mu_0=5log\left(\frac{H_0^{-1}}{Mpc}\right)+25$. The luminosity distance $D_L$ is defined by $D_L(z)=(1+z) H_0 \int \frac{1}{H(z^*)} dz^*$. The $H(z)$ series is limited to the tenth term and approximately integrates the limited series to obtain the luminosity distance.

\subsection{BAO Dataset}
Early Universe is being studied by analyzing BAO. There are three types of BAO measurements namely: High-resolution Sloan Digital Sky Surveys (SDSS), Six Degree Field Galaxy Surveys (6dFGS), and BOSS \cite{Percival_2010_401}. We present results from SDSS, 6dFGS, and BOSS-DR12 based on available BAO data. The following expressions for measurable quantities are used to obtain BAO constraints:

\begin{eqnarray}
    d_A (z)&=&\int_{0}^{z} \frac{dz^*}{H(z^*)}, \label{19}\\
    D_v (z)&=&\left( \frac{d_A (z)^2 z}{H(z)}\right)^{\frac{1}{3}}, \label{20}
\end{eqnarray}
and
\begin{eqnarray} \label{21}
    \chi^2_{BAO}=X^T C^{-1} X,
\end{eqnarray}
where the angular diameter distance and the dilation scale are represented by $d_A(z)$, $D_V(z)$, respectively, and $C$ represents the covariance matrix \cite{Giostri_2012_2012_027}.

\begin{figure*} [!htb]
\centering
\includegraphics[width=7.78cm,height=5.5cm]{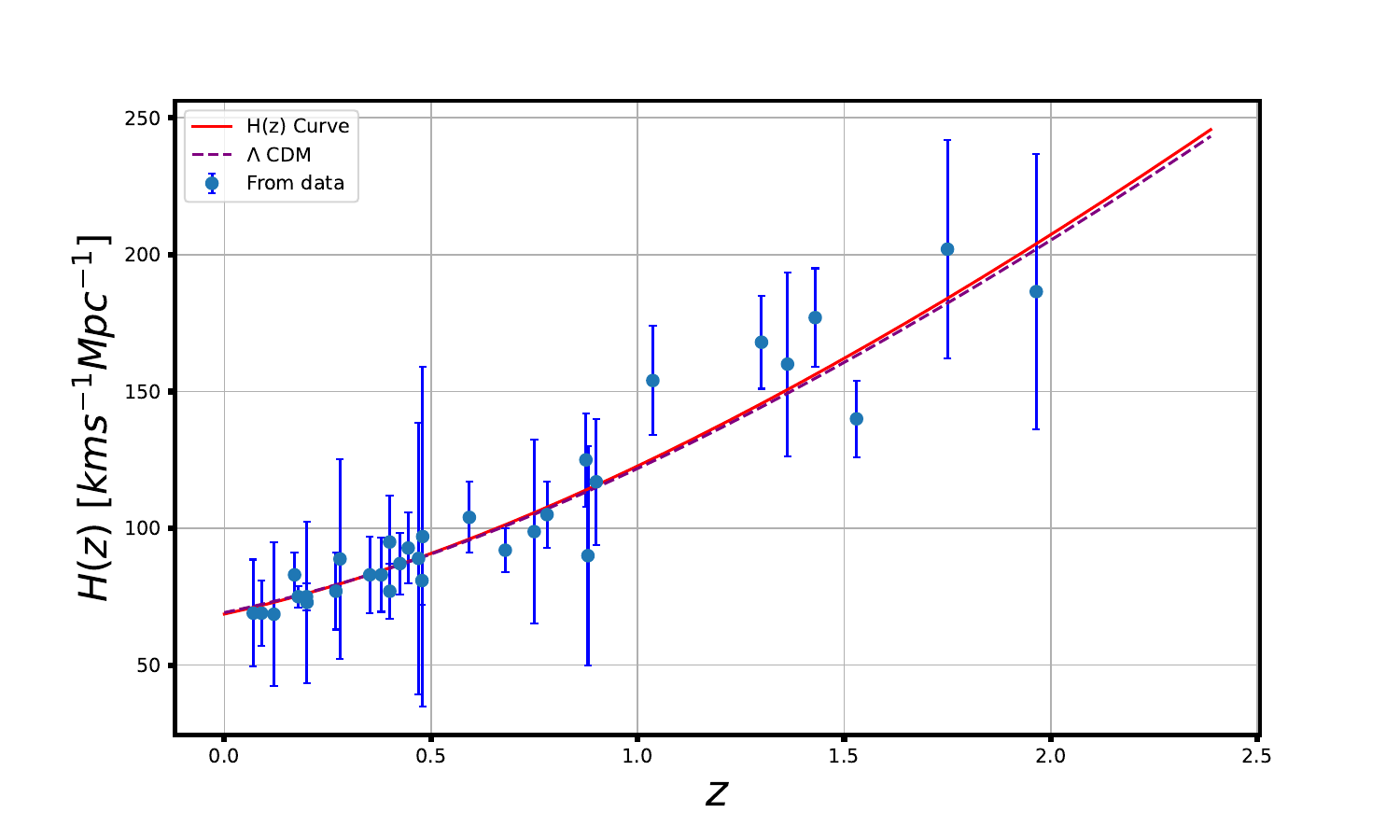}
\includegraphics[width=7.78cm,height=5.5cm]{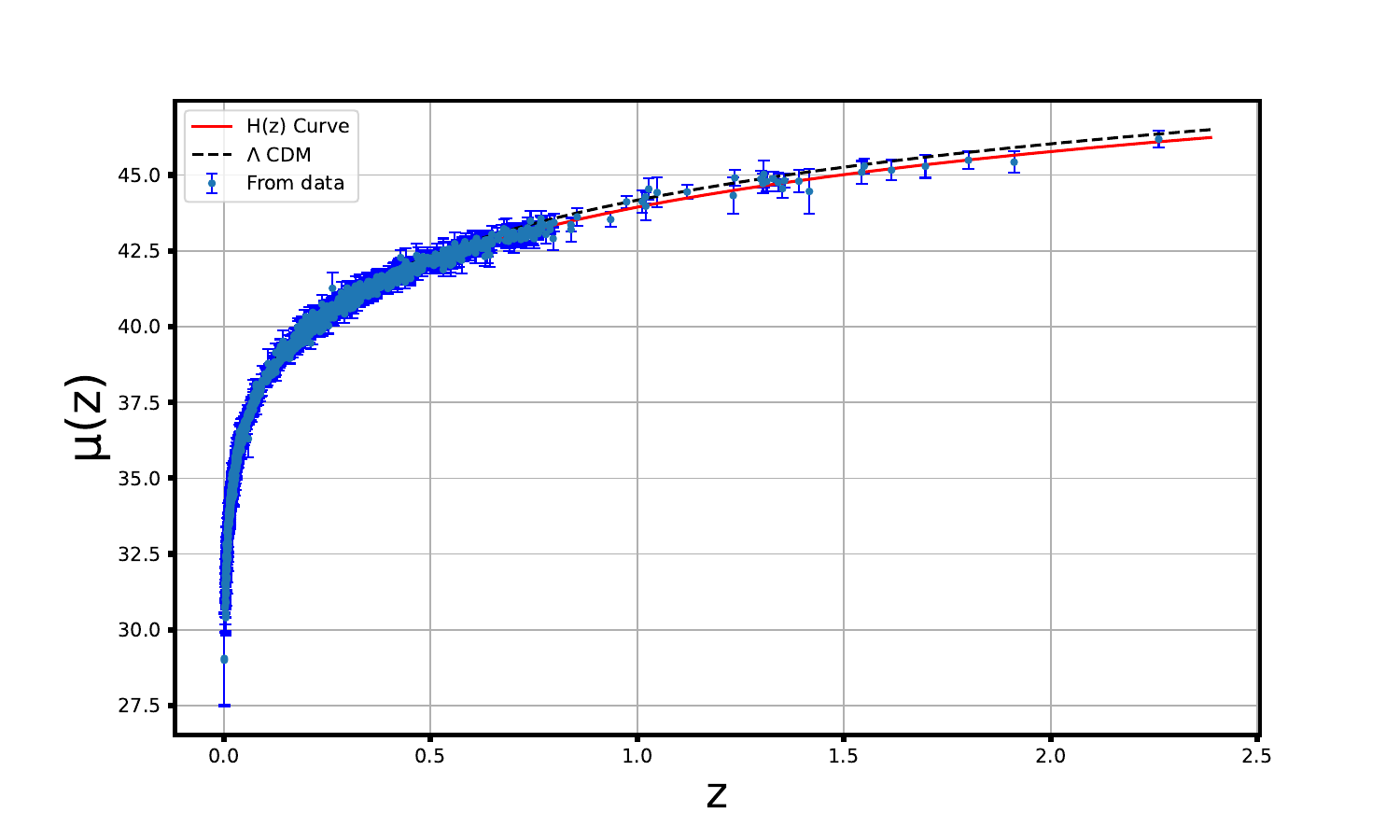}
\caption{The blue error bars are from the 32 points of the CC sample, the solid red line is of the model, and the broken black line is for the $\Lambda$CDM (left panel). In (right panel), the red line is the plot of the model's distance modulus $\mu(z)$ versus $z$, which exhibits a better fit to the 1701 points of the \textit{Pantheon$^+$} data sets along with its error bars.}
\label{FIG1}
\end{figure*}
\begin{figure*} [!htb]
\centering
\includegraphics[scale=0.44]{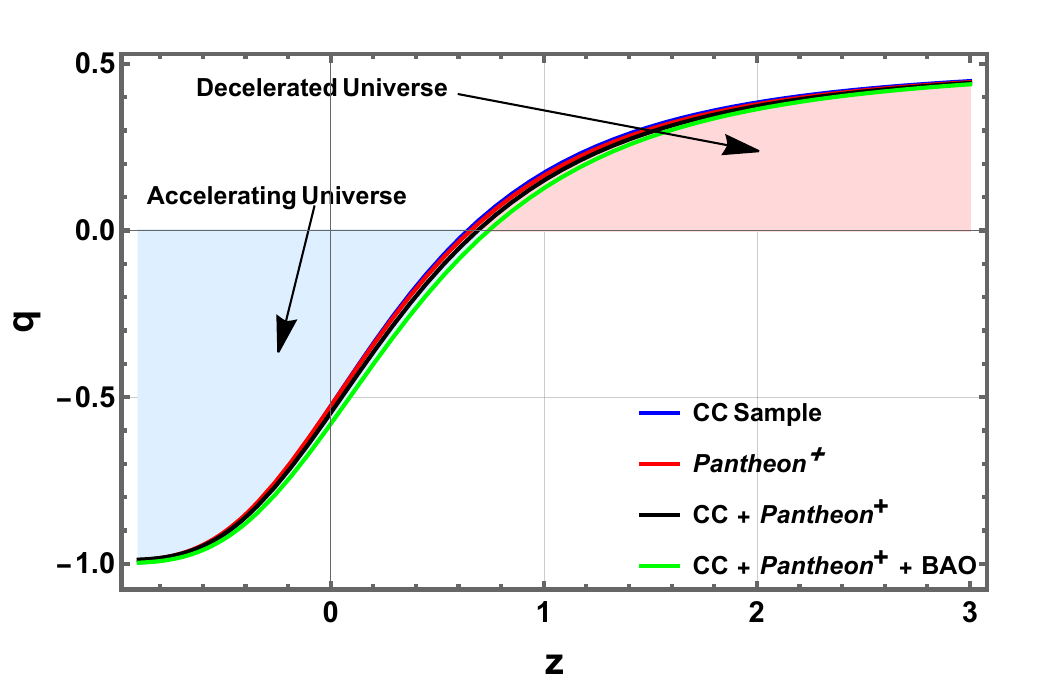}
\includegraphics[scale=0.44]{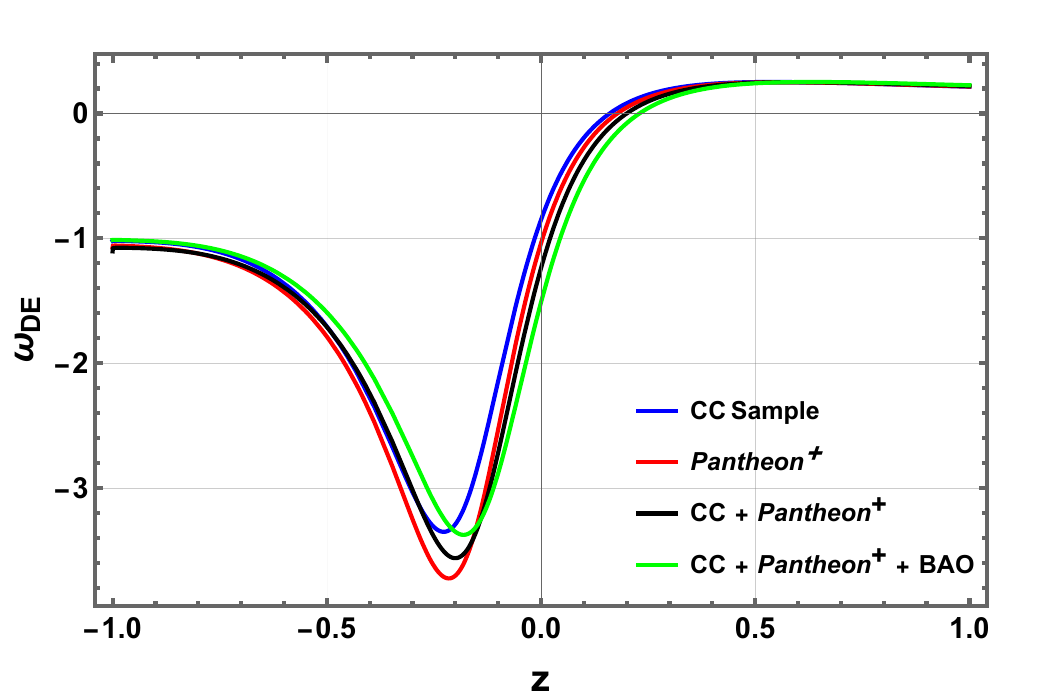}
\caption{Deceleration parameter and EoS parameter with CC, \textit{Pantheon$^+$} and BAO datasets for the parameters $\alpha=1.1$, $\beta=4$.}
\label{FIG4}
\end{figure*}

\begin{table*}[!htb]
\small\addtolength{\tabcolsep}{-3pt}
\centering 
\begin{tabular}{|*{5}{c|}}\hline
\parbox[c][0.7cm]{2cm}{\textbf{Parameters}} & \textbf{CC Sample} & \textbf{\textit{Pantheon$^+$}} & \textbf{CC + \textit{Pantheon$^+$}} & \textbf{CC + \textit{Pantheon$^+$} + BAO}\\ [0.5ex]
\hline \hline
\parbox[c][0.7cm]{2cm}{$\hspace{.85cm} q$} & -0.526 ($z_t \approx 0.636$) & -0.529 ($z_t \approx 0.656$) & -0.548 ($z_t \approx 0.691$) & -0.579 ($z_t \approx 0.74$) \\
\hline
\parbox[c][0.7cm]{2cm}{\hspace{.7cm}$\omega_{DE}$} & -0.8478 & -1.02 & -1.224 & -1.47 \\
\hline
\parbox[c][0.7cm]{2cm}{\hspace{.7cm}$\omega_{eff}$} & -0.684 & -0.686 & -0.7 & -0.72 \\
\hline
\end{tabular}
\caption{Present value of deceleration and EoS parameters based on the CC samples, \textit{Pantheon$^+$} samples, and BAO datasets.} 
\label{TABLE II}
\end{table*}

The contour plots with $1-\sigma$ and $2-\sigma$ errors are given in figure \ref{FIG2} for the CC and $Pantheon^+$ sample data sets, whereas in figure \ref{FIG3} for the CC$+Pantheon^+$ and CC$+Pantheon^+$+BAO data sets. In figure \ref{FIG1}, one can observe that the $H(z)$ curve is lying well within the error bars. All the values obtained for the parameters are listed in Tables \ref{TABLE I} and \ref{TABLE II}.
The deceleration parameter $q=-1-\frac{\dot{H}}{H^2}$ describes the rate of acceleration of the Universe, where a positive $q$ indicates that the Universe is in a decelerated phase, while a negative $q$ indicates that the Universe is in an accelerated phase. The constrained values of model parameters in the Hubble, \textit{Pantheon$^+$}, and BAO data sets resulted in $q$ changing from a positive value at the past, suggesting an early slowdown, to a negative value at the present, indicating an acceleration at present, as seen in figure \ref{FIG4}. In the current cosmic epoch, Hubble and Pantheon data are relatively consistent with the range $q_0=-0.528^{+0.092}_{-0.088}$ determined by recent observations \cite{Christine_2014_89} and a redshift from deceleration to acceleration occurs at $z_t=0.8596^{+0.2886}_{-0.2722}$, $z_t=0.65^{+0.19}_{-0.01}$ \cite{Yang_2020_2020_059, Capoziello_2008_664, Capozziello_2014_90_044016}. The deceleration parameter $q_0= -0.526$, $q_0= -0.529$, $q_0=-0.548$ and $q_0=-0.579$ at the current cosmic epoch and our derived model shows a smooth transition from a deceleration phase of expansion to an acceleration phase, at $z_t = 0.636$, $z_t = 0.656$, $z_t = 0.691$ and $z_t = 0.74$ for CC, \textit{Pantheon$^+$}, CC + \textit{Pantheon$^+$} and CC + \textit{Pantheon$^+$} + BAO datasets, respectively. The recovered transition redshift value $z_t$ is consistent with certain current constraints based on 11 $H(z)$ observations reported by Busca et al. \cite{Busca_2013_552} between the redshifts $0.2 \leq z \leq 2.3$, $z_t = 0.74 \pm 0.5$ from Farooq et al. \cite{Farooq_2013_766}, $z_t = 0.7679^{+0.1831}_{-0.1829}$ by Capozziello et al. \cite{Capozziello_2014_90_044016} and $z_t = 0.60^{+0.21}_{-0.12}$ by Yang et al. \cite{Yang_2020_2020_059}

Among the parameters that define the behavior of the Universe is the deceleration parameter, which determines whether the Universe continuously decelerates or accelerates constantly, has a single phase of transition or several, etc. Energy sources play a similar role in the evolution of the Universe according to the EoS parameter $\big(\omega_{DE}=\frac{p_{DE}}{\rho_{DE}}\big)$. Calculating the related energy density and pressure of DE, as illustrated in figure \ref{FIG4}, it allows us to see the variations in the effective EoS of DE with respect to the redshift variable. The present value of EoS for dark energy $\omega_{DE}(z = 0)$ respectively obtained as, $-0.8478$, $-1.02$, $-1.224$ and $-1.47$ for CC, \textit{Pantheon$^+$}, CC + \textit{Pantheon$^+$} and CC + \textit{Pantheon$^+$} + BAO datasets. It shows the phantom behavior (at $z \leq -0.015$) and its approach to $-1$ at late times. Whereas the present value of effective EoS ($\omega_{eff}$) parameter respectively obtain as $-0.684, \, -0.686, \, -0.7, \, -0.72$ [Table \ref{TABLE II}]. The numerical value of the EoS parameter has also been restricted by several cosmological investigations, including the Supernovae Cosmology Project $\omega_{DE}=-1.035^{+0.055}_{-0.059}$ \cite{Amanullah_2010_716}, Planck 2018, $\omega_{DE}=-1.03\pm 0.03$ \cite{Aghanim_2018_641} and WAMP+CMB, $\omega_{DE}=-1.079^{+0.090}_{-0.089}$ \cite{Hinshaw_2013_208}.

\section{Dynamical System Analysis} \label{SEC-IV}
The $F(R, \mathcal{G})$ gravity model has been able to address some of the key issues of the early and late Universe, and it is always good to know its general phase space structure. Among higher-order theories of gravity, $F(R, \mathcal{G})$ gravity has one of the most complicated field equations, and dynamical system analysis has been important in understanding its physical behavior. Often, the dynamical systems analysis is performed to find the stability of the model and the presence of fixed points \cite{Santos_da_Costa_2018_35,Olmo_2005_72}. In addition, this may also help in avoiding the challenges of solving nonlinear cosmological equations. This will allow us to examine the asymptotic behavior of cosmological models. Analyzing the asymptotic behavior of critical points of the dynamical system, the overall dynamic of the Universe in terms of cosmological epochs can be described. Further from the phase-space analysis, one can assess the stability of the critical points. The dynamical system analysis approach has been used in the modified generalised cosmological models as in Refs. \cite{Dent_2011_009, Ivanov_2012_18, Mirza_2017_011, Bahamonde_2019_100_8, Duchaniya_2023_83, Duchaniya_2022_82, Narawade_2022_36, Kadam_2022_82, Odintsov_2018_98, Odintsov_2019_36, Odintsov_2017_96}.

The dynamical system analysis for the future behavior of the system can predict cosmological models based on dynamical systems. There may be an equation of the type $x= f(x)$ representing the dynamical system, where $x$ represents the column vector, and $f(x)$ represents the equivalent column vector of the autonomous equations. The prime represents the derivative with respect to $N = ln a$. This method can generate the general form of the dynamical system for the modified FLRW equations, which are defined by equation (\ref{8}). As an autonomous system, the set of cosmological equations of the model is written with the following dimensionless variables \cite{Santos_da_Costa_2018_35}:
\begin{eqnarray} \label{22}
    u_1=\frac{\dot{F}_R}{H F_R},\hspace{0.5cm} u_2=\frac{F}{6 H^2 F_R},\hspace{0.5cm} u_3=\frac{R}{6 H^2},\nonumber\\ u_4=\frac{\mathcal{G} F_{\mathcal{G}}}{6 H^2 F_R},\hspace{0.5cm} u_5=\frac{4 H \dot{F}_\mathcal{G}}{F_R},
\end{eqnarray}
with the energy density parameters
\begin{eqnarray} \label{23}
    u_6=\Omega_r= \frac{\kappa^2 \rho_r}{3 H^2 F_R},\hspace{1cm} u_7=\Omega_m= \frac{\kappa^2 \rho_m}{3 H^2 F_R}
\end{eqnarray}

Thus, we have the algebraic identity
\begin{equation} \label{24}
    1=-u_1-u_2+u_3+u_4-u_5+\Omega_r+\Omega_m\\
\end{equation}

The dynamical system is
\begin{eqnarray}
    \frac{du_1}{dN}&=& \frac{\ddot{F}_{R}}{F_{R}H^{2}} -u_1^{2}-u_1\frac{\dot{H}}{H^{2}}, \label{25}\\
    \frac{du_2}{dN}&=& \frac{\dot{F}}{6F_{R}H^{3}}-u_1u_2-2u_2\frac{\dot{H}}{H^{2}}, \label{26}\\
    \frac{du_3}{dN}&=& \frac{\dot{R}}{6H^{3}}-2u_3 \frac{\dot{H}}{H^{2}}, \label{27}\\
    \frac{du_4}{dN}&=& \frac{\dot{\mathcal{G}}}{\mathcal{G}H} u_4+\frac{\mathcal{G}}{24H^{4}} u_5-u_1u_4-2u_4(u_3-2), \label{28}\\
    \frac{du_5}{dN}&=& u_5 \frac{\dot{H}}{H^{2}}+4\frac{\ddot{F_{\mathcal{G}}}}{F_{R}}-u_1u_5, \label{29}\\
    \frac{du_6}{dN}&=& -2u_3 u_6-u_1 u_6, \label{30}\\
    \frac{du_7}{dN}&=& -u_7 \left(3+u_1+2 \frac{\dot{H}}{H^{2}}\right). \label{31}
\end{eqnarray}

To close the system, all terms on the right-hand side of the above equations must be expressed in terms of variables specified in equation (\ref{14}). Thus, we find
\begin{eqnarray}
 \frac{\dot{H}}{H^{2}}&=&u_3-2, \label{32}\\
 \frac{\dot{F}}{6F_{R}H^{3}}&=&{-u_1 u_3}, \label{33}\\
 \frac{\dot{R}}{6H^{3}}&=&{u_1 u_3}, \label{34}\\
 \frac{\mathcal{G}}{24H^{4}}&=&u_3-1, \label{35}\\
 \frac{\dot{\mathcal{G}}}{\mathcal{G}H}&=&\frac{1}{u_3-1}\left[{u_1 u_3}+2(u_3-2)^{2}\right]. \label{36}
\end{eqnarray}

\begin{table*} [!htb]
    \centering 
    \begin{tabular}{|*{6}{c|}}\hline
    \parbox[c][0.8cm]{0.85cm}{\textbf{C.P.}} & \textbf{$u_3$} & \textbf{$u_4$} & \textbf{$u_6$} & \textbf{$u_7$} & \textbf{Exists for} \\ [0.5ex] 
    \hline\hline 
    \parbox[c][0.8cm]{0.85cm}{$\mathcal{P}_{1}$} & 0 & 0 & 1 & 0 & always \\
     \hline
    \parbox[c][0.8cm]{0.85cm}{$\mathcal{P}_{2}$} & 0 & $\frac{1-u_6}{5}$ & $u_6$ & 0 & $3 + 2 u_6 \neq 0$, $\beta=\frac{1}{2}$ \\
     \hline
    \parbox[c][0.8cm]{0.85cm}{$\mathcal{P}_{3}$} & 2 & -2 & 0 & 0 & $-1+4\beta \neq 0$ \\
   \hline
    \parbox[c][0.8cm]{0.85cm}{$\mathcal{P}_{4}$} & 0 & $u_4$ & 0 & 0 & $-1+u_4 \neq 0, -1+2 u_4 \neq 0, \beta =\frac{-3-u_4}{8(-1+u_4)}$ \\
    \hline
    \parbox[c][0.8cm]{0.85cm}{$\mathcal{P}_{5}$} & $u_3$ & $\frac{1}{2}(-6+u_3)$ & 0 & 0 & $-1+u_3 \neq 0, -2+u_3 \neq 0, 14-12 u_3+3 u_3^2 \neq 0, \beta =0$ \\
    \hline
    \end{tabular}
     \caption{The critical points of the dynamical system. The coordinates of the critical points: ($u_3, u_4, u_6, u_7$).}
    \label{TABLE-III}
\end{table*}

A theory specified by $\Gamma=\frac{\ddot{F}_{R}}{F_{R}H^{2}}$ is used. It can be inferred that the system can only be considered complete once it is expressed in terms of dynamical variables (\ref{22}), (\ref{23}). From equations (\ref{14}), (\ref{22}) and equation (\ref{29}), we can get
\begin{eqnarray}
    u_3&=&2u_2, \label{37}\\
    u_5&=&\frac{u_4}{u_3-1}\left[{2u_1}+\frac{\beta-1}{u_3-1}\left[2(u_3-3)^{2}+{u_1u_3}\right]\right]. \label{38}
\end{eqnarray}

Using these relations and the constraint [equation (\ref{24})], the system can be reduced to a set of four equations as
\begin{eqnarray}
    \frac{du_3}{dN}&=&u_1u_3-2u_3(u_3-2), \label{39}\\
    \frac{du_4}{dN}&=&\frac{\beta u_4}{u_3-1}\left[2(u_3-3)^{2}+{u_1u_3}\right]+u_1u_4-2u_4(u_3-2),\nonumber\\ \label{40}\\
    \frac{du_6}{dN}&=&-2u_3 u_6-u_1 u_6, \label{41}\\
    \frac{du_7}{dN}&=&-u_7(2u_3+u_1-1), \label{42}
\end{eqnarray}
where
\begin{equation} \label{43}
    u_1 = \frac{-1+\frac{3}{2}u_3+u_6+u_7+u_4-2(\beta-1)\frac{(u_3-2)^{2}}{(u_3-1)^{2}}u_4}{1+\frac{u_4}{(u_3-1)}\left[2+u_3\frac{(\beta-1)}{(u_3-1)}\right]},
\end{equation}
and
\begin{eqnarray} \label{omega_total}
    \omega_{eff}=-1-\frac{2\dot{H}}{3H^2}=-1-\frac{2}{3}u_3.
\end{eqnarray}

\begin{table} [H]
    \centering 
    \begin{tabular}{|*{6}{c|}}\hline
    \parbox[c][0.6cm]{0.85cm}{\bf{C.P.}} & \bf{$\Omega_m$} & \bf{$\Omega_r$} & \bf{$\Omega_{de}$} & \bf{$q$} & \bf{$\omega_{eff}$} \\ [0.5ex] 
    \hline\hline 
    \parbox[c][0.6cm]{0.5cm}{$\mathcal{P}_{1}$} & 0 & 1 & 0 & 1 & $\frac{1}{3}$ \\
     \hline
    \parbox[c][0.6cm]{0.5cm}{$\mathcal{P}_{2}$} & 0 & $u_6$ & $1-u_6$ & 1 & $\frac{1}{3}$ \\
     \hline
    \parbox[c][0.6cm]{0.5cm}{$\mathcal{P}_{3}$} & 0 & 0 & 1 & -1 & $-1$ \\
    \hline
    \parbox[c][0.6cm]{0.5cm}{$\mathcal{P}_{4}$} & 0 & 0 & 1 & 1 & $\frac{1}{3}$ \\
    \hline
    \parbox[c][0.6cm]{0.5cm}{$\mathcal{P}_{5}$} & 0 & 0 & 1 & $1-u_3$ & $\frac{1}{3} (1-2u_3)$ \\
    \hline
    \end{tabular}
     \caption{The deceleration, EoS and density parameters for the critical points.} 
    \label{TABLE-IV}
\end{table}

\begin{table} [H]
\centering
\begin{tabular}{|*{2}{c|}}\hline
\parbox[c][0.5cm]{0.85cm}{\textbf{C.P.}} & \parbox[c][1cm]{12cm}{\textbf{Eigenvalues}}\\\hline \hline
\parbox[c][0.5cm]{0.5cm}{$\mathcal{P}_1$} & \parbox[c][1cm]{12cm}{$\big\{4, -1, 1, -4(-1+2 \beta)\big\}$}\\\hline
\parbox[c][0.5cm]{0.5cm}{$\mathcal{P}_2$} & \parbox[c][1cm]{12cm}{$\big\{0, \frac{-5 (-1+2 u_6)}{3+2 u_6}, 1, 4\big\}$}\\\hline
\parbox[c][0.5cm]{0.5cm}{$\mathcal{P}_3$} & \parbox[c][1cm]{12cm}{$\big\{-4, -3, \frac{3-12\beta-\sqrt{9-136 \beta +400 \beta^2}}{2(-1+4 \beta)}, \frac{3-12 \beta +\sqrt{9-136 \beta +400 \beta^2}}{2(-1+4 \beta)}\big\}$}\\\hline
\parbox[c][0.5cm]{0.5cm}{$\mathcal{P}_4$} & \parbox[c][1cm]{12cm}{$\big\{\frac{4u_4}{(-1+u_4)(-1+2 u_4)}, \frac{-3+u_4}{-1+u_4}, \frac{2(-1+3 u_4)}{-1+u_4}, \frac{1-7 u_4+10 u_4^2}{(-1+u_4)(-1+2 u_4)} \big\}$}\\\hline
\parbox[c][0.5cm]{0.5cm}{$\mathcal{P}_5$} & \parbox[c][1cm]{12cm}{$\big\{0, \frac{-6(1-2 u_3+u_3^2)}{14-12 u_3+3 u_3^2}, -4(-1+u_3), (5-4u_3) \big\}$}\\\hline
\end{tabular}
\caption{Equivalent eigenvalues for fixed points.}
\label{TABLE-VI}
\end{table}

Tables \ref{TABLE-III} and \ref{TABLE-IV} show the conditions under which the critical points of these systems exist and the eigenvalues of these systems. The critical points can be calculated to analyze their features and behavior. Table \ref{TABLE-IV} represents the cosmological parameters correspond to the critical points. Below we will discuss the properties of each critical point and their potential connection with different evolutionary eras of the Universe, which are divided into five critical points.

\subsection{Visualization of Phase Portraits}
For a complete understanding of the distinguishing features of each critical point, it is crucial to describe its behavior in proper diagrams. The phase portraits for each critical point are presented in this section, along with the critical steps involved in their derivation and whether they are compatible with the analysis of Tables \ref{TABLE-III} and \ref{TABLE-VI}. The properties of each of the five critical points are separately discussed, and explore their possible connections to the eras of the evolution of the Universe that they represent.

\begin{itemize}
    \item {\bf{Point $\mathcal{P}_1$:}} In a radiation-dominated Universe, the first critical point $\mathcal{P}_1$ occurs. Table \ref{TABLE-III} shows that the critical point exists for all values of the free parameters. This critical point applies to any free model parameter based on Table \ref{TABLE-IV}, $\Omega_r=1$. The EoS parameter  $\omega_{tot}=\frac{1}{3}$ and deceleration parameter $q=1$ demonstrate that the background level does not experience late-time acceleration in this solution. Table \ref{TABLE-III} shows that our critical point is a saddle hyperbolic. Point $\mathcal{P}_1$ possesses a 2D local unstable manifold with boundaries defined only within the neighbourhood of the critical point, whereas the description of local indicates that these boundaries are determined only within the neighbourhood of the critical point.
\end{itemize}

\begin{figure*} [!htb]
    \centering
    \includegraphics[width=58mm]{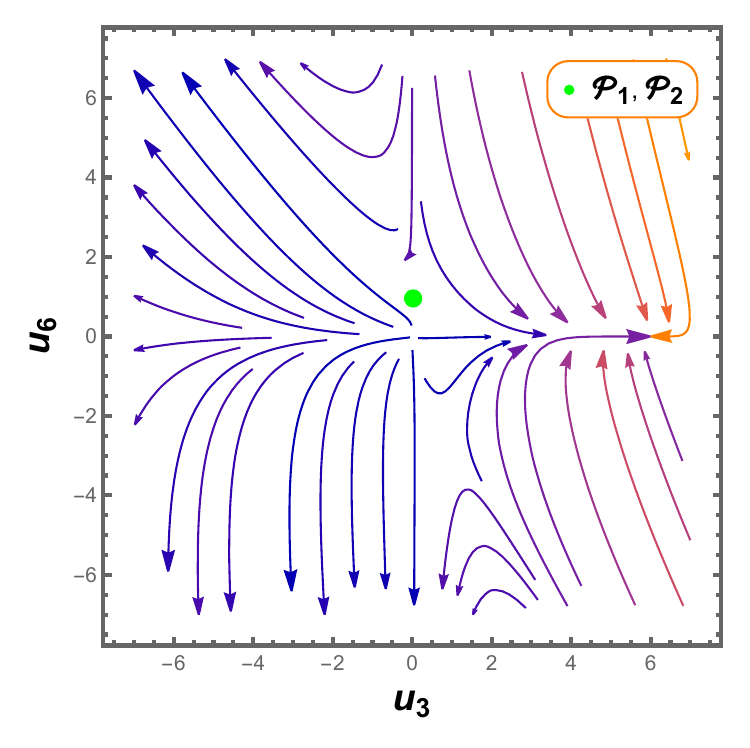}
    \includegraphics[width=58mm]{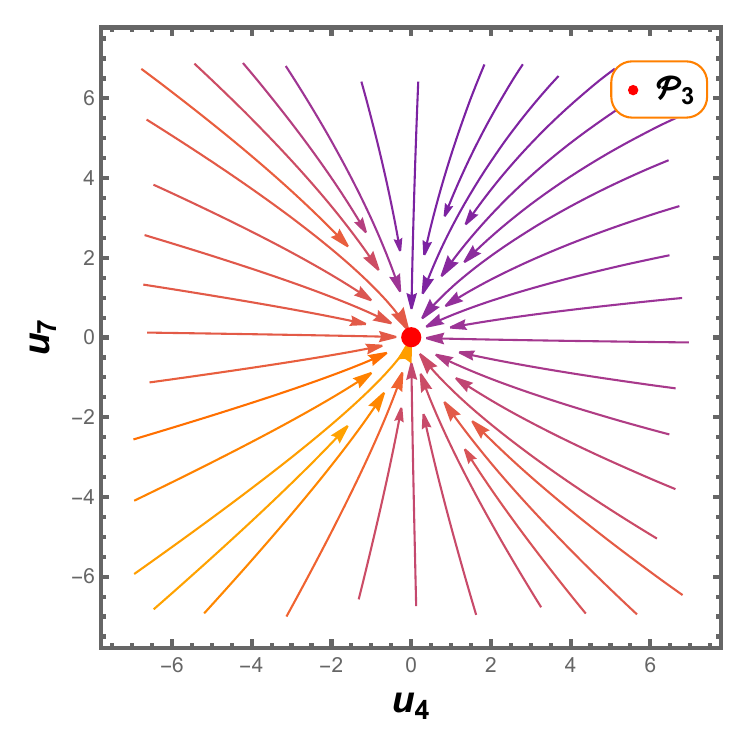}\\
    \includegraphics[width=58mm]{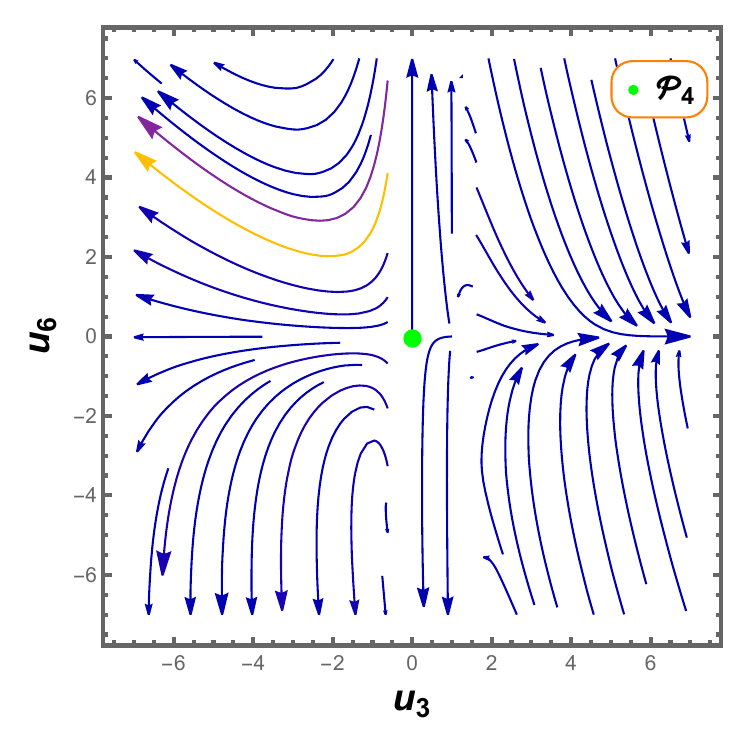}
    \includegraphics[width=58mm]{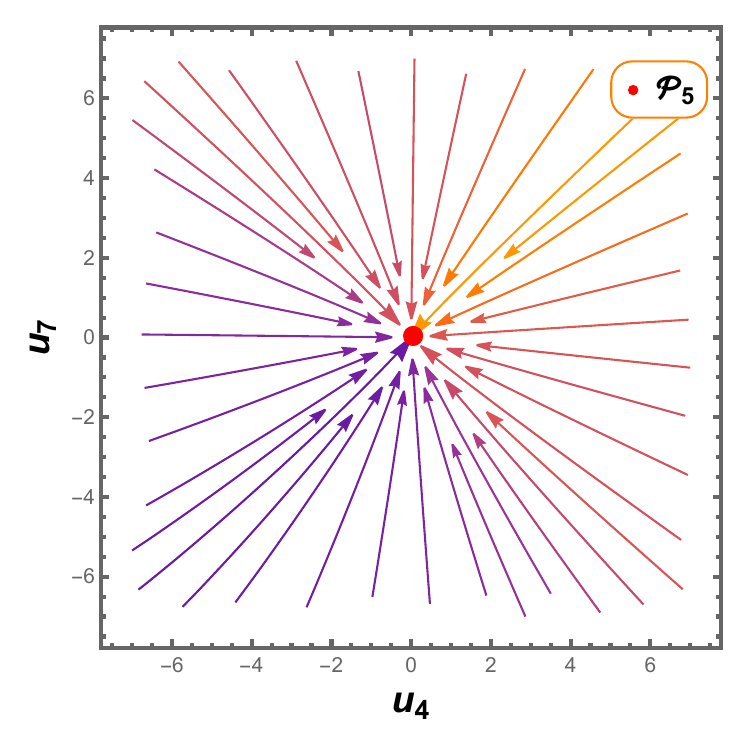}
    \caption{$2D$ phase portrait for the dynamical system.} \label{FIG5}
\end{figure*}
\begin{table*} [!htb]
    \centering 
    \begin{tabular}{|*{4}{c|}}\hline
    \parbox[c][0.6cm]{0.75cm}{\textbf{C.P.}} & \textbf{Acceleration equation} & \textbf{Phase of the Universe} & \textbf{Stability condition} \\ [0.5ex] 
    \hline\hline 
    \parbox[c][0.6cm]{0.5cm}{$\mathcal{P}_{1}$} & $\dot{H}=-2 H^2$ & $a(t)= t_{0} (2 t+c_{1})^\frac{1}{2}$ & Unstable  \\
     \hline
    \parbox[c][0.6cm]{0.5cm}{$\mathcal{P}_{2}$} & $\dot{H}=-2 H^2$ & $a(t)= t_{0} (2 t+c_{1})^\frac{1}{2}$ & Unstable  \\
     \hline
    \parbox[c][0.6cm]{0.5cm}{$\mathcal{P}_{3}$} & $\dot{H}=0$ & $a(t)=t_0 e^{c_1 t}$ & Stable \\
   \hline
    \parbox[c][0.6cm]{0.5cm}{$\mathcal{P}_{4}$} & $\dot{H}=-2 H^2$ & $a(t)= t_{0} (2 t+c_{1})^\frac{1}{2}$ & Unstable \\
    \hline
    \parbox[c][0.6cm]{0.5cm}{$\mathcal{P}_{5}$} & $\dot{H}=(-2+u_3) H^2$ & $a(t)=t_0 \left((2-u_3)t+c_1\right)^\frac{1}{2-u_3}$ & Stable \\
    \hline
    \end{tabular}
     \caption{Phase of the Universe with stability conditions.} 
    \label{TABLE-V}
\end{table*}
\begin{itemize}
    \item {\textbf{Point $\mathcal{P}_2$:}} Table \ref{TABLE-III} shows that the second critical point $\mathcal{P}_2$ exists for $3 + 2 u_6 \neq 0$ and $\beta=\frac{1}{2}$. The Universe is in a radiation-dominated phase with $\Omega_r=u_6$, $\Omega_m=0$, and $\Omega_{DE}=1-u_6$. This is further evidenced by the EoS parameter ($\omega_{tot}$) being equal to $\frac{1}{3}$ and the deceleration parameter $q$ having a value of 1. The Jacobian matrices associated with these critical points have real positive and negative parts and zero eigenvalues, indicating that it has an unstable saddle behavior.\\
    \item {\textbf{Point $\mathcal{P}_3$:}} Under the conditions in Table \ref{TABLE-III}, this point $\mathcal{P}_3$ corresponds to a Universe dominated by DE. Since it is stable under the conditions shown, it can be considered a late-time phase of the Universe. Interestingly, under conditions with $-1+4 \beta \neq 0$, the EoS parameter ($\omega_{tot}$) equals the value of the cosmological constant $-1$ at this critical point, where $\Omega_{DE} = 1$, $\omega_{tot} = -1$. The deceleration parameter $q = -1$. Since these features are compatible with observations, they are a great advantage of the scenario under consideration; furthermore, they can only be obtained by using $F(R,\mathcal{G})$ gravity without explicitly including a cosmological constant or a canonical or phantom scalar field. It is stable when $0<\beta\leq \frac{9}{100}$. The corresponding eigenvalue is\\ $\big\{-4, -3, \frac{3-12\beta-\sqrt{9-136 \beta +400 \beta^2}}{2(-1+4 \beta)}, \frac{3-12 \beta +\sqrt{9-136 \beta +400 \beta^2}}{2(-1+4 \beta)}\big\}$.\\
    \item {\textbf{Point $\mathcal{P}_4$:}} This critical point exists in a radiation-dominated Universe for $-1 + u_4 \neq 0$, $-1 + 2 u_4 \neq 0$ and $\beta =\frac{-3 - u_4}{8(-1 + u_4)}$, leading to a decelerating phase of the Universe with an EoS parameter $\omega_{tot} =\frac{1}{3}$ and deceleration parameter $q = 1$. The corresponding density parameters are $\Omega_r=0$, $\Omega_r=0$, and $\Omega_r=1$. The eigenvalues associated with this critical point reveal positive and negative signs by taking some restrictions on $u_4$, indicating that it is an unstable node.\\
    \item {\textbf{Point $\mathcal{P}_5$:}} At late times, point $\mathcal{P}_5$ could attract the Universe due to its stability under the conditions presented in Table \ref{TABLE-III}. There are similarities between this point and $\mathcal{P}_3$, but there are differences in parameter regions. In particular, it suggests an accelerating Universe dominated by DE. The negative value of the deceleration parameter indicates the accelerating phase of the Universe, and $\omega_{tot} = -1$ behaves as a cosmological constant at this critical point, where $\Omega_{DE} = 1$ and $q = -1$ at the background. The deceleration parameter shows the accelerating behavior when $1<u_3$. It is stable when $u_3 > \frac{5}{4}$. The corresponding eigenvalue is\\ $\big\{0, \frac{- 6(1 - 2 u_3 + u_3^2)}{14 - 12 u_3 + 3 u_3^2}, -4(-1 + u_3), (5 - 4 u_3) \big\}$.
\end{itemize}

Table \ref{TABLE-V} summarises all the results and the corresponding scale factor.

\begin{figure}[H]
    \centering
    \includegraphics[width=80mm]{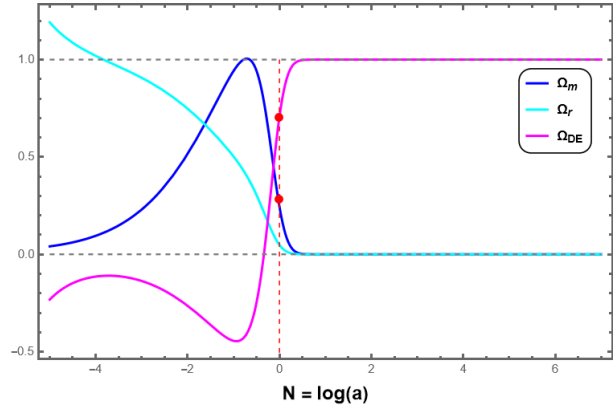}
    \caption{Evolution of density parameters DE (magenta line), matter (blue line) and radiation (cyan line) for the initial conditions: $u_3=10^{-9.45}$, $u_4=0.01$, $u_6=1.28999$, $u_7=0.448 \times 10^{-1.2}$.} 
    \label{FIG6}
\end{figure}

Figure \ref{FIG6}  shows the cosmic evolution of the density parameter for matter, radiation, and DE for the model (\ref{14}) with the initial conditions $u_3=10^{-9.45}$, $u_4=0.01$, $u_6=1.28999$ and $u_7=0.448 \times 10^{-1.2}$. The behavior is consistent with recent cosmic observations on the evolution of density parameters. To obtain the current densities, $\Omega_m \approx 0.28$, $\Omega_{DE} \approx 0.679$, and $\Omega_r \approx 0.047$ are calculated. Radiation dominance is shown in figure \ref{FIG6} at the beginning, followed by a brief phase of matter dominance and, at the end, the de-Sitter phase.

\section{Conclusion} \label{SEC-V}

We have investigated the cosmological behavior of a modified Gauss-Bonnet gravity model class by describing the gravitational action involving the Ricci scalar and Gauss-Bonnet invariant. The parameterization of the Hubble and other geometrical parameters are presented, and the coefficients are constrained using CC sample, the largest \textit{Pantheon$^+$} and the BAO datasets. The best-fit values of the coefficients are obtained and provided in Table \ref{TABLE I}. Subsequently, the deceleration parameter and the EoS parameter are constrained, and the best-fit values are presented in Table \ref{TABLE II}. The model presented here shows a smooth transition from a decelerating phase to an accelerated expansion phase. For CC, \textit{Pantheon$^+$}, CC + \textit{Pantheon$^+$} and CC + \textit{Pantheon$^+$} + BAO data, the transition redshifts are obtained as, $z_t=0.636$, $z_t=0.656$, $z_t=0.691$ and $z_t=0.74$, respectively. The dark energy EoS parameter shows that the expansion of the Universe has increased since it is within the phantom region for $z \leq -0.015$. The DE EoS parameter at $z = 0$ is $-0.8478$, $-1.02$, $-1.224$ and $-1.47$ for CC, \textit{Pantheon$^+$}, CC + \textit{Pantheon$^+$} and CC + \textit{Pantheon$^+$} + BAO datasets respectively, which are within the recent findings of cosmological observations. Also, the present value of effective EoS ($\omega_{eff}$) parameter obtained to be, $-0.684, \, -0.686, \, -0.7, \, -0.72$ respectively for CC, \textit{Pantheon$^+$}, CC + \textit{Pantheon$^+$} and CC + \textit{Pantheon$^+$} + BAO datasets. 

In the second phase of the analysis, we have studied the dynamical system analysis focusing on the type of $F(R, \mathcal{G})$ function considered. This allows us to analyze global behavior and stability in terms of the cosmological model. We found some interesting preliminary findings for the finite phase space of a power-law class of fourth-order gravity models $\mathcal{F}(R, \mathcal{G})=\alpha R^2 \mathcal{G}^\beta$. Equations (\ref{39})-(\ref{42}) present the dynamical system for the mixed power law $\mathcal{F}(R, \mathcal{G})$ gravity model. Table \ref{TABLE-III} provides critical points and existing conditions for the model. At the same time, Table \ref{TABLE-IV} presents a value for the deceleration, EoS, and density parameters. A total of five critical points are obtained, two of which ($\mathcal{P}_3, \mathcal{P}_5$) are stable and three ($\mathcal{P}_1, \mathcal{P}_2, \mathcal{P}_4$) are unstable. During the de-Sitter phase of the Universe, stable critical points appeared, whereas unstable behavior was observed during the radiation-dominated phase. A signature of eigenvalues and a phase-space portrait support the behavior of critical points. The trajectory behavior indicates that the unstable critical points act as release points while the stable ones act as attractor points (figure \ref{FIG5}).

The accelerating behavior of the model has been confirmed from the value of EoS ($\omega_{tot}=-1$) and deceleration parameter ($q=-1$). The density parameters are obtained to be $\Omega_m \approx 0.28$, $\Omega_{DE} \approx 0.679$, and $\Omega_r \approx 0.047$. Figure \ref{FIG6} shows radiation dominance, followed by matter dominance, and finally, de-Sitter dominance. In the first phase, using the cosmological data sets, we have obtained the present value of matter density and DE density parameters respectively as $\Omega_m\approx 0.28$ and $\Omega_{DE}\approx 0.68$. From the evolution plot of the dynamical system analysis that uses the initial conditions for the dynamical variables, we have shown that the present value of the matter density and DE density parameters, respectively, $\approx 0.28$ and $\approx 0.679$. We conclude that in the both approaches, the value of the density parameters are aligned, and the stable accelerating behavior of the model has been confirmed.

\begin{appendix}
\section*{Appendix} 
\begin{table}[H]\label{Appendix-I}
\small\addtolength{\tabcolsep}{-2pt}
\centering 
\begin{tabular}{c c c c c | c c c c c | c c c c c} 
\hline 
No. & Redshift & $H(z)$ & $\sigma_{H(z)}$ & Ref. & No. & Redshift & $H(z)$ & $\sigma_{H(z)}$ & Ref. &  No. & Redshift & $H(z)$ & $\sigma_{H(z)}$ & Ref.\\ [0.5ex] 
\hline 
1. & 0.070 & 69.00 & 19.6 & \cite{Zhang_2014_14} & 12. & 0.400 & 95.00 & 17.00 &  \cite{Simon_2005_71} & 23. & 0.875 & 125.00 & 17.00 &  \cite{Moresco_2012_2012_006}\\ 
2. & 0.090 & 69.00 & 12.0 & \cite{Simon_2005_71} & 13. & 0.4004 & 77.00 & 10.20 &  \cite{Moresco_2016_2016_014} & 24. & 0.880 & 90.00 & 40.00 & \cite{Stern_2010_2010_008}\\
3. & 0.120 & 68.60 & 26.2 & \cite{Zhang_2014_14} & 14. & 0.425 & 87.10 & 11.20 &  \cite{Moresco_2016_2016_014}  & 25. & 0.900 & 117.00 & 23.00 &  \cite{Simon_2005_71}\\
4. & 0.170 & 83.00 & 8.00 &  \cite{Simon_2005_71}  & 15. & 0.445 & 92.80 & 12.90 &  \cite{Moresco_2016_2016_014}  & 26. & 1.037 & 154.0 & 20.00 &  \cite{Moresco_2012_2012_006}\\
5. & 0.179 & 75.00 & 4.00 & \cite{Moresco_2012_2012_006} & 16. & 0.47 & 89.00 & 49.60 &  \cite{Ratsimbazafy_2017_467}  & 27. & 1.300 & 168.0 & 17.00 &  \cite{Simon_2005_71} \\
6. & 0.199 & 75.00 & 5.00 &  \cite{Moresco_2012_2012_006} & 17. & 0.4783 & 80.90 & 9.00 &  \cite{Moresco_2016_2016_014}  & 28. & 1.363 & 160.00 & 33.60 &  \cite{Moresco_2015_450} \\
7. & 0.200 & 72.90 & 29.60 &  \cite{Zhang_2014_14} & 18. & 0.48 & 97.00 & 62.00 &  \cite{Stern_2010_2010_008}  & 29. & 1.430 & 177.0 & 18.00 &  \cite{Simon_2005_71} \\ 
8. & 0.270 & 77.00 & 14.00 &  \cite{Simon_2005_71}  & 19. & 0.593 & 104.00 & 13.00 &  \cite{Moresco_2012_2012_006} & 30. & 1.530 & 140.0 & 14.00 &  \cite{Simon_2005_71}\\
9. & 0.280 & 88.80 & 36.60 & \cite{Zhang_2014_14} & 20. & 0.680 & 92.00 & 8.00 &  \cite{Moresco_2012_2012_006} & 31. & 1.750 & 202.0 & 40.00 & \cite{Simon_2005_71}\\ 
10. & 0.352 & 83.00 & 14.00 &  \cite{Moresco_2012_2012_006}& 21. & 0.750 & 98.80 & 33.60 &  \cite{Borghi_2022_928} & 32. & 1.965 & 186.5 & 50.4 & \cite{Simon_2005_71}\\
11. & 0.380 & 83.00 & 13.50 &  \cite{Moresco_2016_2016_014} & 22. & 0.781 & 105.00 & 12.00 & \cite{Moresco_2012_2012_006} & &  &  &  & \\[0.5ex] 
\hline 
\end{tabular} 
\caption{$H(z)$ measurements were made using the CC technique, expressed in [km s$^{-1}$ Mpc$^{-1}$] units, along with the corresponding errors.} 
\label{table: Table II} 
\end{table}
\end{appendix}
\section*{Acknowledgement} SVL acknowledges the financial support provided by University Grants Commission (UGC) through Senior Research Fellowship  (UGC Ref. No. 191620116597) to carry out the research work. BM acknowledges the support of IUCAA, Pune (India), through the visiting associateship program. The authors are thankful to the honourable referees for their valuable comments and suggestions for the improvement of the manuscript.

\section*{References}
\bibliographystyle{utphys}
\bibliography{ref_short}

\end{document}